\documentclass[preprint,showpacs,showkeys,preprintnumbers,amsmath,amssymb]{revtex4}
 \usepackage{dcolumn}
 \usepackage{bm}
 \usepackage{graphicx}
 \usepackage{subfigure}
 \begin{document}

 \title{Time evolutions of scalar field perturbations in $D$-dimensional Reissner-Nordstr\"om Anti-de Sitter black holes}

 \author{Ran Li$^1$}
 \thanks{Email: liran@htu.edu.cn}

 \author{Hongbao Zhang$^{2,3}$}
 \thanks{Email: hzhang@vub.ac.be}

 \author{Junkun Zhao$^1$}
 \thanks{Email: zhaojkun1991@163.com}

 \affiliation{
 $^1$ Department of Physics, Henan Normal University, Xinxiang 453007, China\\
 $^2$ Department of Physics, Beijing Normal University, Beijing 100875, China\\
 $^3$ Theoretische Natuurkunde, Vrije Universiteit Brussel, and The International Solvay Institutes,
 Pleinlaan 2, B-1050 Brussels, Belgium}

 \begin{abstract}

 Reissner-Nordstr\"om Anti-de Sitter (RNAdS) black holes are
 unstable against the charged scalar field perturbations due to the well-known
 superradiance phenomenon. We present the time domain analysis of charged
 scalar field perturbations in the RNAdS black hole background in general dimensions.
 We show that the instabilities of charged scalar field can be explicitly
 illustrated from the time profiles of evolving scalar field. By using
 the Prony method to fit the time evolution data, we confirm the mode
 that dominates the long time behavior of scalar field is in accordance with
 the quasinormal mode from the frequency domain analysis. The superradiance origin
 of the instability can also be demonstrated by comparing the real part of the dominant mode
 with the superradiant condition of charged scalar field. It is shown that all
 the unstable modes are superradiant, which is consistent with the analytical result in the frequency domain analysis. Furthermore, we also confirm there exists the rapid exponential growing modes in the RNAdS case, which makes the RNAdS black hole a good test ground to investigate the nonlinear evolution of superradiant instability.

 \end{abstract}



 \maketitle

 \section{introduction}

 The studying of black holes in the asymptotically AdS spacetime is mainly
 motivated by the gauge/gravity duality \cite{ads}.
 This novel duality provides us a powerful tool to study the strongly coupled
 gauge theory by mapping it to weakly coupled classical gravitational system
 living in one higher dimension. In this holographic setup, asymptotically AdS black hole is dual to a thermal state of conformal field theory on its boundary \cite{plya,witten}.
 The construction of various holographic models of condensed matter physics depends highly on the properties of this kind of backgrounds. So understanding the various properties of asymptotically AdS black hole is a central topic in current fundamental research.

 The interaction of charged field with the electromagnetic background in a charged black hole
 is more interesting than expected because of the superradiance phenomenon. This effect is mainly known for the rotating black hole, for which it allows a bosonic wave to carry out the rotating energy of the black hole \cite{zeldovich,bardeen,misner,starobinsky}. However, the charged scalar field in the vicinity of charged black hole can also extract the electromagnetic energy of the black hole by a wave with increased amplitude via superradiant scattering process \cite{bekenstein}. If, in addition, the scattering wave is reflected by a potential barrier or a mirror sufficiently far away, then the wave will be amplified repeatedly. The resulted exponential growth is dubbed as the superradiant instability, or black hole bomb \cite{press,cardoso2004bomb}. Because of the theoretical significance and astrophysical application of superradiant instability, it has attracted a lot of attention in recent years \cite{Rosa,leejhep,jgrosa,hod2013prd,hodbhb,strafuss,dolan,Hod,hodPLB2012,konoplyaPLB,DiasPRD2006,
   cardoso2004ads,cardoso2006prd,KKZ,aliev,uchikata,rlplb,knopolya,rlepjc,randilaton,
   degolladoprd2013,degollado,raduprl,liepjc2015,dolanprd2013,binwang,
   liepjc2014,liplb2015,liprd,ranliplbtime2015,Ishibashi,wangmengjie2015,daojunliu}. one can refer to Ref.\cite{superradiance} for a comprehensive review on this topic.

 In the RNAdS black hole case, the timelike boundary of AdS black hole is equivalent to the infinite high potential barrier, which implies the absence of dissipation at infinity. So, a charged scalar field in RNAdS black hole background will exhibit superradiant instability, which has been confirmed by calculating the quasinormal modes of RNAdS black hole in $4$-dimension \cite{uchikata} and in general $D$-dimensions \cite{mengjiewang} in the limit of small $r_H/L$ ($r_H$ is the horizon radius of black hole and $L$ is the AdS curvature radius). These results are valid when ignoring the backreaction on the spacetime.
 The understanding of dynamics of charged scalar field in RNAdS black hole requires the full
 nonlinear machinery of Einstein equation, complemented with the linearized analysis. Actually
 the spherically symmetric nonlinear evolution of superradiant instability for the charged scalar field has recently been reported and the corresponding final states have also been studied \cite{bosch}. The static black hole solutions which is
 conjectured to be the end point of the superradiant instability
 were also constructed in \cite{basu,santos} previously. 

 However, the linear time domain analysis of charged scalar field in RNAdS black hole backgrounds in general $D$ dimensions is still missing. This paper intends to fill this gap, which is supposed to be a good preparation for one to attack the more generic nonlinear dynamics of instability. Especially, we also confirm there exists the rapid exponential growing mode when the charge of scalar field is large enough, which makes the RNAdS black hole a good test ground to investigate the nonlinear development of superradiant instability.

 This paper is organized as follows. In Sec. II, we introduce the basic setup of a charged massless scalar field in RNAdS black hole in general $D$-dimensions, and derive the linear evolution equation of the perturbation. In Sec. III, we describe our numerical schemes including the pseudospectral method, Runge-Kutta method, Prony method, and the direct integral method in detail. We present our numerical results and relevant discussions in Sec. IV. Final remarks are made in the last Section.

 \section{framework}

 The background spacetime considered in the present paper is the spherically symmetric Reissner-Nordstr\"om Anti-de Sitter black hole in general $D=n+2$ dimensions, and the metric is given by
 \begin{eqnarray}
 ds^2=-f(r)dt^2+\frac{1}{f(r)}dr^2+r^2 d\Omega_n^2\;,
 \end{eqnarray}
 where $d\Omega_n^2$ is the metric on the unit $n$-sphere, and the metric function $f(r)$ takes
 the form
 \begin{eqnarray}
 f(r)=1-\frac{\mu}{r^{n-1}}+\frac{q^2}{r^{2(n-1)}}+\frac{r^2}{L^2}\;.
 \end{eqnarray}
 The parameters $\mu$ and $q$ are related with the mass and charge of the black hole, and
 $L$ is the AdS radius, which is related to the cosmological constant $\Lambda$ by the formula $L^2=-n(n+1)/2\Lambda$.

 This line element (1) describes the black hole if $f(r)=0$ has two real roots, which correspond to the event horizon $r_+$ and the Cauchy horizon $r_-$, respectively. In addition, the electric charge $q$ of black hole is also constrained by the relation
 \begin{eqnarray}
 q\leq r_+^{n-1}\sqrt{1+\frac{n+1}{n-1}\frac{r_+^2}{L^2}}\equiv q_c\;,
 \end{eqnarray}
 to ensure the positive definiteness of the Hawking temperature.

 The electromagnetic gauge field $A_\mu$ of RNAdS black hole is given by
 \begin{eqnarray}
 A=\left(-\sqrt{\frac{n}{2(n-1)}}\frac{q}{r^{n-1}}+C\right)dt\;,
 \end{eqnarray}
 where the constant $C$ can be gauge fixed. It is argued that this constant just shifts the real part
 of the quasinormal mode without affecting the imaginary part. Therefore, when studying the superradiant instability below, we can take $C=0$ for convenience.

 Note also that all physical quantities can be normalized by the AdS scale $L$. So we shall set the AdS radius $L=1$ in the following numerical calcualtions.

 It turns out to be very useful to adopt the ingoing Eddington-Finkelstein coordinates in studying
 the time domain evolution of the perturbation field, which is defined by
 \begin{eqnarray}
 v=t+r^*\;,
 \end{eqnarray}
 with $r^*$ being the tortoise coordinate, defined as $dr^*=\frac{dr}{f}$.
 Then the metric in the ingoing Eddington-Finkelstein coordinates reads as
 \begin{eqnarray}
 ds^2=-fdv^2+2dvdr+r^2 d\Omega_n^2\;,
 \end{eqnarray}
 and the electromagnetic gauge field in the new coordinates can be written as
 \begin{eqnarray}
 A=-\sqrt{\frac{n}{2(n-1)}}\frac{q}{r^{n-1}}dv\;,
 \end{eqnarray}
 where we have also performed a gauge transformation to eliminate the space component of gauge field.

 For simiplicity, it is sufficient to consider the test scalar field to be
 charged and massless. the dynamics
 is then governed by the Klein-Gordon equation
 \begin{eqnarray}
 (\nabla_\nu-ieA_\nu)(\nabla^\nu-ieA^\nu)\Psi=0\;,
 \end{eqnarray}
 where $e$ denotes the charge of the scalar field.

 By taking the ansatz of the scalar field
 \begin{eqnarray}
 \Psi=\frac{1}{r^{n/2}}\phi(v, r) Y(\theta_i)\;,
 \end{eqnarray}
 where $Y(\theta_i)$ is a scalar spherical harmonic on the $n$-sphere,
 the Klein-Gordon equation is shown to be separable
 in this background. After some algebra, we can finally
 get the following partial differential equation
 \begin{eqnarray} \label{KG}
 &&2\partial_v\partial_r \phi+f\partial^2_r \phi
 +\frac{df}{dr}\partial_r \phi-2ieA_v \partial_r \phi
  -\frac{n}{2}\frac{df}{dr} \frac{\phi}{r}
  \nonumber\\&&
  -ie \frac{dA_v}{dr}\phi
 -\frac{n}{2}\left(\frac{n}{2}-1\right)\frac{f}{r^2}\phi
 -\lambda\frac{\phi}{r^2}=0\;,
 \end{eqnarray}
 where $\lambda=l(l+n-1)$ is the $n$-dimensional spherical harmonic eigenvalue.
 We can now numerically evolve the scalar field according to this partial differential equation once the initial condition is given.

 \section{Numerical Method}

 \subsection{Time domain analysis}
 In the space direction, we introduce a new radial coordinate $z=r_+/r$ to transform the spatial
 domain $[r_+, +\infty]$ to a compact domain $[0, 1]$. It is obvious that the spatial infinity of AdS
 is precisely located at $z=0$ in the new coordinate. So we can set the boundary condition at the spatial infinity of AdS as $\phi(v, z=0)=0$. We then expand the scalar field in a basis of Chebyshev polynomials in the space direction. The scalar field function is discretized by those values on the Gauss-Lobatto collocation points. The spatial derivative of the scalar field can be computed by multiplying the Gauss-Lobatto derivative matrix. One can
 refer to Ref.\cite{santosreview,hongbaoreview} for a recent review on
 this method to discretize the space direction, which is usually named as pseudospectral method.

 Next we use the forth order Runge-Kutta scheme to integrate the scalar field in time direction. To be more precise, we first evolve the space derivative of the scalar field according to (\ref{KG}) and then integrate out the scalar field by multiplying the integral matrix. It should be noted that the integral matrix was constructed according to the aforementioned boundary condition on the boundary of RNAdS black hole, which is automatically maintained in the whole evolution process.
 The initial condition of the scalar field configuration will be chosen as the Gaussian wave packet as usual. It is noteworthy that Our method to evolve the perturbation field, namely pseudo-spectral method in the spacial direction supplemented by Runge-Kutta method in temporal direction, is very different from the usually used finite difference scheme.

 In order to extract the quasinormal mode that is dominant in the evolving process, we use the
 Prony method of fitting the time domain profile data by superposition of damping or growing
 exponents
 \begin{eqnarray}
 \phi(v, r) \simeq \sum_{i=1}^{p}C_i e^{-i\omega_i(v-v_0)}\;.
 \end{eqnarray}
 We consider the late time behavior of the perturbation, which starts at $v_0$ and ends at $t=\mathcal{K}\Delta v+v_0$, where $\mathcal{K}$ is a integer and $\mathcal{K}\geq 2p-1$.
 Then the above formula is valid for each value from the profile data
 \begin{eqnarray}
 x_{k}\equiv\phi(k\Delta v+v_0, r)=\sum_{i=1}^{p}C_i e^{-i\omega_ik\Delta v}=
 \sum_{i=1}^{p}C_i z_i^k\;.
 \end{eqnarray}
 The Prony method allows us to find $z_i$ in terms of the known $x_k$, and to
 calculate out the quasinormal modes $\omega_i$ since $\Delta v$ is also known.
 The dominant mode is then selected as the one that has the greatest amplitude $C_i$
 among the $p$ modes we obtained.

 \subsection{Frequency domain analysis}

 For the sake of comparing our time domain analysis with the frequency domain analysis,
 we shall use the shooting method to calculate the quasinormal mode. Firstly,
 we take the Fourier transformation of our evolution equation (\ref{KG}) to derive
 the eigenvalue equation for the quasinormal modes. Then by taking the regular boundary
 condition near the event horizon (Note that the ingoing boundary condition is automatically
 achieved by this regular boundary condition in the ingoing Eddington-Finkelstein coordinates), we can integrate the eigenvalue equation from a point $r_s$ very close the horizon outwards up to a radial value $r_m$.
 Similarly, we can also integrate the eigenvalue equation inward from infinity by taking the fall off boundary condition, where the infinity may be taken as $r_l=10^3 r_+$.
 The obtained two solution at the intermediate radius $r_m$ should satisfy the condition that
 their Wronskian is equal to zero. Then this condition can be used to solve the quasinormal mode
 in terms of a secant method. We take different values of $r_m$ to check the numerical accuracy.
 We also compare our numerical results for quasinormal modes with those reported in Ref.\cite{uchikata} and
 \cite{mengjiewang}, and find that they are in good agreement.

 \section{result}

\begin{figure}
\subfigure{\includegraphics[width=3in]{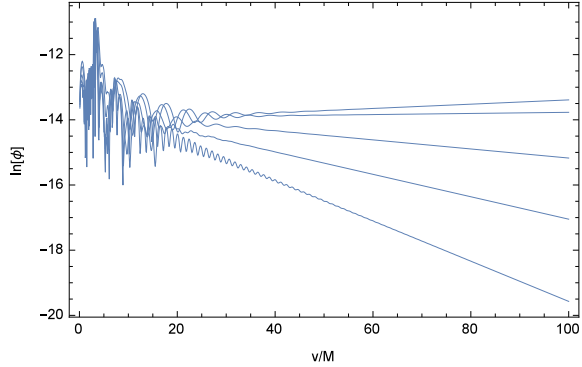}}
\caption{Time evolution of charged massless scalar field with the initial Gaussian wave packet in the 4-dimensional RNAdS black hole. The parameters are taken as $r_+=0.1, l=0$, and $e=4$. This panel shows the logarithm of scalar field evolving with the time $v$. From the bottom to top, $q/q_c$ is taken as
0.2, 0.4, 0.6, 0.8, and 0.9, respectively.}
\end{figure}

 \begin{table}
 \caption{Quasinormal modes for the 4-dimensional RNAdS black hole
 with $r_+=0.1, l=0$, and $e=4$. The first column gives the values of $q/q_c$, where
 $q_c=0.101489$ in the present case. The second column is taken
 from the time domain analysis by fitting the evolution data using the
 Prony method, while the third column is calculated from the frequency domain analysis
 by using the direct integration method.}
\centering
\begin{tabular}{lccc}
\hline
\hline
$q/q_c$ & $\omega_T$ & $\omega_F$ \\ \hline
$0.2$\;\; &  $2.84146701-0.061698331i$\;\;   &$2.84146698-0.061697901i$ \\
$0.4$\;\; &  $2.96501483-0.034407650i$\;\;   &$2.96501505-0.034408063i$ \\
$0.6$\;\; &  $3.06720935-0.013945787i$\;\;   &$3.06720959-0.013945621i$ \\
$0.8$\;\; &  $3.15154864+0.001731455i$\;\;   &$3.15154882+0.001731427i$ \\
$0.9$\;\; &  $3.18786861+0.006574396i$\;\;   &$3.18786870+0.006574302i$ \\
 \hline
 \hline
\end{tabular}
\end{table}

 Firstly, we consider the time evolution of charged massless scalar field in the 4-dimensional
 RNAdS black hole. The time profiles of the evolving scalar field are plotted in Fig.(1),
 from which we can explicitly observe that the the perturbations will exponentially decay with small $q/q_c$, and grow when $q/q_c$ is big. This shows that the small RNAdS black hole is always stable against the charged scalar perturbations when $q/q_c$ is small. An analytical results of the superradiant instability condition was derived in Ref.\cite{mengjiewang} by using the asymptotic
 matching method for the general $D$ dimensions case, which is given by the following formula
 \begin{eqnarray}
 \frac{q}{q_c}> \sqrt{\frac{2(n-1)}{n}}\frac{2N+n+l+1}{e}\;,
 \end{eqnarray}
 with an integer $N$ $(N\geq 0)$.
 In our present cases, the condition ($N=0$ gives the most strict condition on the instability) implies the scalar field will be unstable in the parameter range of $q/q_c>0.75$.
 The corresponding modes that dominant the time evolutions and the quasinormal modes from the frequency domain analysis are presented in Table (1). Obviously, the modes from the time domain analysis show good agreement with the results from the frequency domain analysis . In addition, we have checked that our numerical results are also consistent with those in Ref.\cite{uchikata}. The superradiance condition (in the case of 4-dimensional RNAdS black hole) is given by
 \begin{eqnarray}
 \omega_c=\frac{eq}{r_+}\;.
 \end{eqnarray}
 It can be easily checked that all unstable modes are superradiant, while all the stable modes are not superradiant. This reveals the superradiant origin of the instability, which is also consistent with the analytical result in frequency domain analysis. These conclusions give us confidence in studying the various cases of scalar field evolution in the following cases.

We then explore the influence of scalar field charge on the time evolution
of the perturbations by taking $e=6$. The time profiles of the evolving scalar field
and the corresponding dominant modes are presented in Fig.(2) and Table II,
respectively.
For the case of $q/q_c=0.9$, we see that the logarithm of scalar field amplitude
increases oscillatorily with the time. In fact, by employing the shooting
method to calculate the corresponding quasinormal modes, we can obtain two superradiant unstable modes. One is listed in the Table II, and the other is given by $\omega=5.07228723+0.011844848i$. The behavior of scalar perturbation is thus governed by the superposition of these two unstable modes. However, one can see the mode that dominates the late time evolution is just the one list in the Table II,
because this mode has a bigger imaginary part than the other one. This conclusion is also confirmed by fitting the evolving data using the Prony method, where the dominant mode is selected as the one that has the greatest amplitude $C_i$. We have also checked that the plot of the late time evolution becomes a
straight line finally.

\begin{figure}
\subfigure{\includegraphics[width=3in]{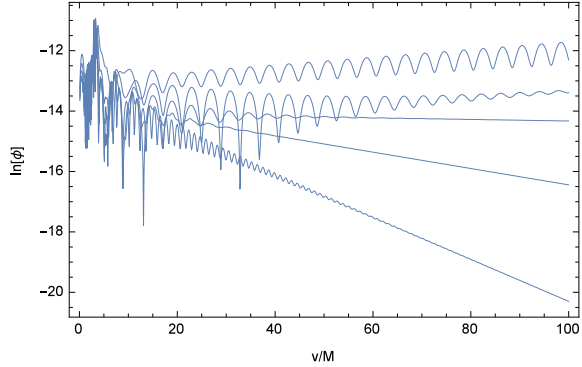}}
\caption{Time evolution of charged massless scalar field in the 4-dimensional RNAdS black hole. The parameters are taken as $r_+=0.1, l=0$, and $e=6$. From the bottom to top, $q/q_c$ is taken as
0.2, 0.4, 0.6, 0.8, and 0.9, respectively.}
\end{figure}

 \begin{table}
 \caption{Dominant quasinormal modes corresponding to the case of Fig.(2).}
\centering
\begin{tabular}{lccc}
\hline
\hline
$q/q_c$ & $\omega_T$ & $\omega_F$ \\ \hline
$0.2$\;\; &  $2.81047285-0.069467051i$\;\;   &$2.81047281-0.069466677i$ \\
$0.4$\;\; &  $3.02131058-0.027084738i$\;\;   &$3.02131083-0.027084408i$ \\
$0.6$\;\; &  $3.20293579-0.002579454i$\;\;   &$3.20293622-0.002579450i$ \\
$0.8$\;\; &  $3.35901781+0.010220341i$\;\;   &$3.35901799+0.010219998i$ \\
$0.9$\;\; &  $3.49061169+0.015754157i$\;\;   &$3.49061162+0.015754123i$ \\
 \hline
 \hline
\end{tabular}
\end{table}

 We also study the time evolution of charged scalar field with $l=1$.
 The corresponding time profiles of the evolving scalar field
 and the dominant modes are presented in Fig.(3) and Table III, respectively.
 Frequency domain analysis shows that the imaginary parts of $l=1$ unstable modes are generically three orders of magnitude less than that of $l=0$ unstable modes. It is known that the growth time scale is set by the inverse of the imaginary part of dominant mode. This implies that the $l=0$ unstable modes have a much faster growth rate than the $l=1$ unstable modes. So, only after a longer time evolution for the $l=1$ case, can a significant growth of scalar field amplitude be observed. To achieve this, we have actually taken the time of evolution $v$ as $2.5\times10^4$.

\begin{figure}
\subfigure{\includegraphics[width=3in]{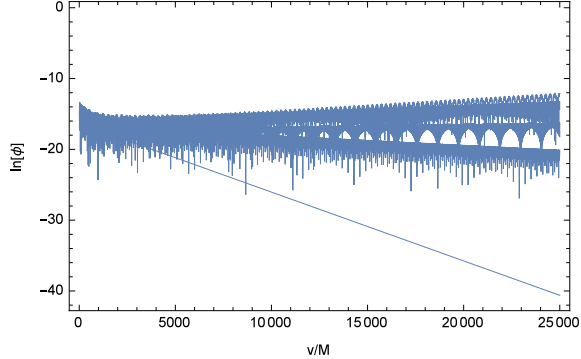}}
\caption{Time evolution of charged massless scalar field in the 4-dimensional RNAdS black hole. The parameters are taken as $r_+=0.1, l=1$, and $e=10$. From the bottom to top, $q/q_c$ is taken as
0.2, 0.4, 0.6, 0.8, and 0.9, respectively.}
\end{figure}

 \begin{table}
 \caption{Dominant quasinormal modes corresponding to the case of Fig.(3).}
\centering
\begin{tabular}{lccc}
\hline
\hline
$q/q_c$ & $\omega_T$ & $\omega_F$ \\ \hline
$0.2$\;\; &  $3.92534630-0.000971835i$\;\;   &$3.92534609-0.000972179i$ \\
$0.4$\;\; &  $4.15457354-0.000138129i$\;\;   &$4.15457354-0.000138503i$ \\
$0.6$\;\; &  $4.35731369+0.000038309i$\;\;   &$4.35731366+0.000037463i$ \\
$0.8$\;\; &  $4.53944906+0.000082312i$\;\;   &$4.53944848+0.000081130i$ \\
$0.9$\;\; &  $4.70458149+0.000166153i$\;\;   &$4.70458055+0.000164961i$ \\
 \hline
 \hline
\end{tabular}
\end{table}

 We also consider the time evolution of $l=1$ scalar field in $D=5$ dimension, which is plotted
 in Fig.(4). The corresponding dominant quasinormal modes are listed in Table IV.
 We take the black hole charge $q/q_c$ fixed and vary the scalar field charge $e$.
 It is observed that the amplitude of scalar field decays with the evolution time for the small scalar charge $e$, which coincides with the observation that the superradiant instability condition (13)
 will not be satisfied for the small scalar charge. It should be noted that the superradiant instability condition (13) is derived under the assumption of $l=p(n-1)$ with $p$ being a non-negative integer.
 But as we see above, the formula (13) is also valid in the current case. It is tempted to conjecture that the superradiant instability condition (13) is universal although the analytical derivation is still absent.

\begin{figure}
\subfigure{\includegraphics[width=3in]{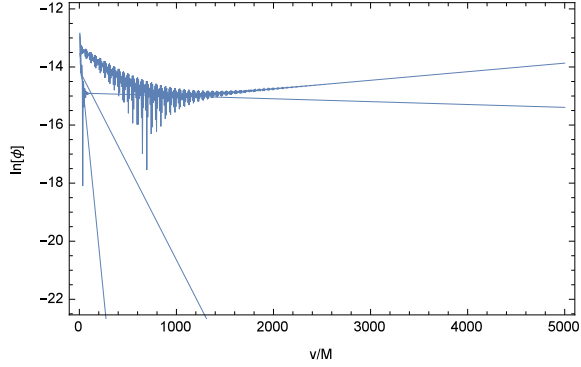}}
\caption{Time evolution of charged massless scalar field in the 5-dimensional RNAdS black hole . The parameters are taken as $r_+=0.2, l=1$, and $q/q_c=0.9$. From the bottom to top, the scalar field charge $e$ is taken as
$2, 4, 6$, and $8$, respectively.}
\end{figure}

 \begin{table}
 \caption{Dominant quasinormal modes corresponding to the case of Fig.(4).}
\centering
\begin{tabular}{lccc}
\hline
\hline
$e$ & $\omega_T$ & $\omega_F$ \\ \hline
$2$\;\; &  $4.82777711-0.033526869i$\;\;   &$4.82777707-0.033526871i$ \\
$4$\;\; &  $4.95526147-0.006531186i$\;\;   &$4.95526150-0.006531165i$ \\
$6$\;\; &  $5.06887892-0.000098657i$\;\;   &$5.06887895-0.000098657i$ \\
$8$\;\; &  $5.16455429+0.000295220i$\;\;   &$5.16455431+0.000295228i$ \\
 \hline
 \hline
\end{tabular}
\end{table}

 The evolutions of charged scalar field in general $D$ dimensional RNAdS black holes
 are also investigated. For the convenience of comparison, the
 time profiles of evolving scalar field in $D=4,5,6$ and $7$ dimensional RNAdS black holes
 are plotted in Fig.(5), while the corresponding dominant quasinormal modes are
 listed in Table V. It is shown that, the charged scalar field has a larger growth rate in lower spacetime dimension by keeping other parameters invariant. Correspondingly, the dominant quasinormal mode has a bigger imaginary part in lower dimension. Also, we can conclude
 that, in the RNAdS black hole with large dimensions, superradiant instability can be generated by
 the scalar field with the large charge.

\begin{figure}
\subfigure{\includegraphics[width=3in]{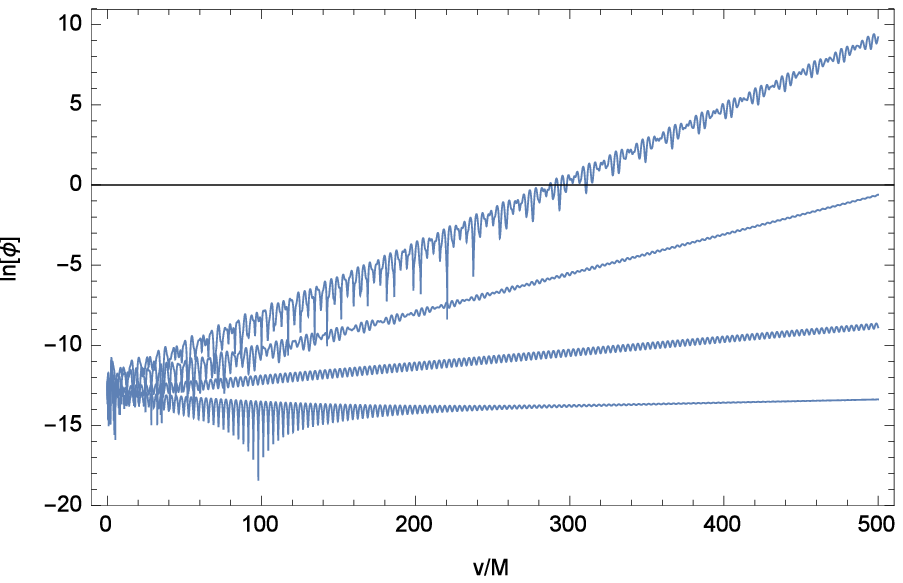}}
\caption{Time evolution of charged massless scalar field in general $D$ dimensions. The parameters are taken as $r_+=0.2, l=0, q/q_c=0.9$, and $e=10$. From the top to bottom, $D$ is taken as
4, 5, 6, and 7, respectively.}
\end{figure}

 \begin{table}
 \caption{Dominant quasinormal modes corresponding to the case of Fig.(5).}
\centering
\begin{tabular}{lccc}
\hline
\hline
$D$ & $\omega_T$ & $\omega_F$ \\ \hline
$4$\;\; &  $6.66110040+0.045019055i$\;\;   &$6.66117717+0.044943392i$ \\
$5$\;\; &  $6.37313389+0.024546532i$\;\;   &$6.37313908+0.024529405i$ \\
$6$\;\; &  $7.06680375+0.008350309i$\;\;   &$7.06680355+0.008350474i$ \\
$7$\;\; &  $6.02811697+0.002004873i$\;\;   &$6.02811698+0.002004864i$ \\
 \hline
 \hline
\end{tabular}
\end{table}

It is shown in the work \cite{degolladoprd2013,degollado,liplb2015, ranliplbtime2015} that there exists rapid growth unstable modes in the charged Reissner-Nordstrom black hole and the charged string black hole. Motivated by this observation, we would like to end this section by demonstrating a very rapidly growing time evolution in Reissner-Nordstrom AdS black hole in four and five dimensions. Figure (6) displays the time evolutions of charged scalar field in these two cases.
The two panels show that the amplitudes of scalar field grow rapidly in
the very short evolution time scale.
It should be noted that, when the charge of scalar field becomes too big, the
frequency domain analysis of quasinormal modes becomes difficult because the equation becomes stiff.
This makes the comparison of the results from the time domain and the frequency domain analysis much harder.
The dominant quasinormal modes are presented in Table VI.
For the current case, the modes from the two sides are not matched as precisely as the previous cases.

\begin{figure}
\subfigure{\includegraphics[width=3in]{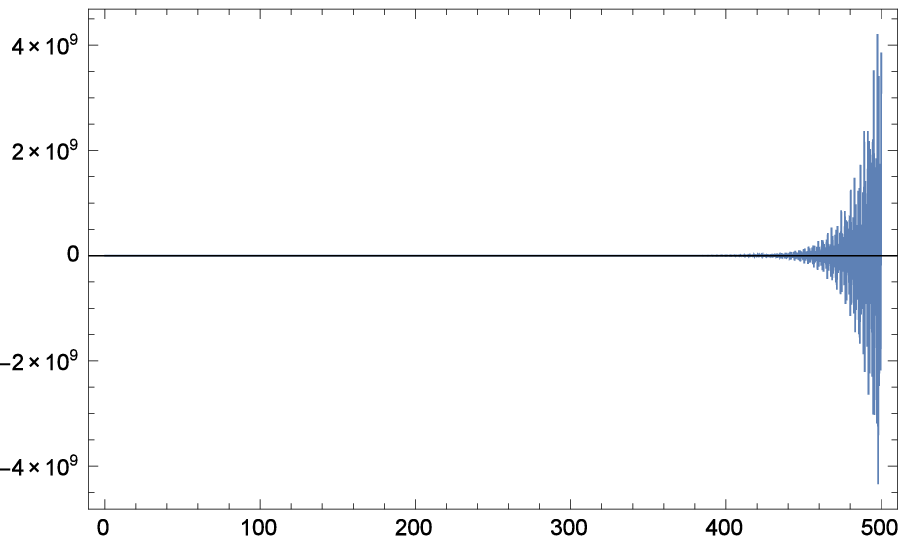}}
\subfigure{\includegraphics[width=3in]{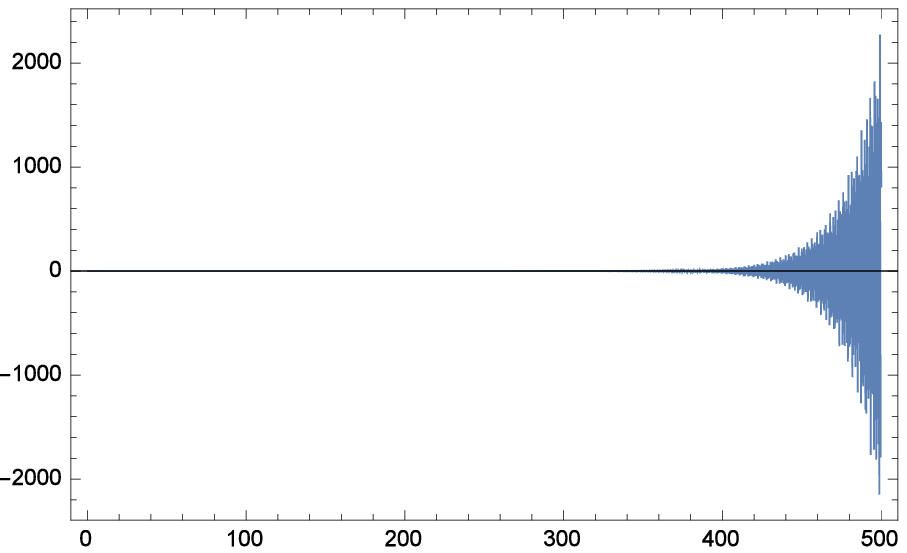}}
\caption{The time profiles of scalar field amplitudes in $4$ and $5$ dimensions. The parameters are taken as $r_+=0.1, l=0, q/q_c=0.9$. The scalar field charge takes $e=30$ for the 4-dimensional case,
and $e=40$ for the 5-dimensional case. }
\end{figure}

 \begin{table}
 \caption{Dominant quasinormal modes corresponding to the case of Fig.(6).}
\centering
\begin{tabular}{lccc}
\hline
\hline
$D$ & $\omega_T$ & $\omega_F$ \\ \hline
$4$\;\; &  $8.05661084+0.072314791i$\;\;   &$8.05628772+0.072555923i$ \\
$5$\;\; &  $8.94700037+0.040946808i$\;\;   &$8.94408885+0.042057342i$ \\
 \hline
 \hline
\end{tabular}
\end{table}

 \section{Conclusion}

 In this paper, we have studied the long time evolution of charged massless scalar field perturbation in the background of Reissner-Nordstr\"om Anti-de Sitter black hole in general $D$ dimensions. We have adopted pseudo-spectral method and the forth-order Runge-Kutta method to evolve the scalar field with the initial Gaussian wave packet. It is shown by our numerical scheme that, by fitting the evolution data using the Prony method, the obtained dominant mode coincides well with the unstable modes computed from the frequency domain analysis. We also confirmed the existence of the rapid growing modes of scalar field with large charge, which is important for the nonlinear evolution of charged superradiant instability.

 \section*{ACKNOWLEDGEMENT}

  R. L. and J. Z. are supported by NSFC with Grant No. 11205048. R. L. is also supported by the Foundation for Young Key Teacher of Henan Normal University. H. Z. is supported in part by the Belgian Federal Science Policy Office through the Interuniversity Attraction Pole P7/37, by FWO-Vlaanderen through the project G020714N, and by the Vrije Universiteit Brussel through the Strategic Research Program "High-Energy Physics". He is also an individual FWO fellow supported by 12G3515N.

 \end{document}